\documentclass[12pt]{article}

\usepackage{amssymb,amsmath}

\newcommand{\be}{\begin{equation}}
\newcommand{\ee}{\end{equation}}
\newcommand{\Dlt}{\Delta}
\newcommand{\dlt}{\delta}
\newcommand{\prt}{\partial}

\newcommand{\bt}{\beta}

\newcommand{\ep}{\varepsilon}

\newcommand{\ra}{\rightarrow}

\newcommand{\om}{\omega}
\newcommand{\Om}{\Omega}

\newcommand{\dgr}{\dagger}
\newcommand{\lbd}{\lambda}

\newcommand{\cF}{{\cal F}}

\newcommand{\cA}{{\cal A}}

\newcommand{\rgl}{\rangle}
\newcommand{\lgl}{\langle}

\begin{document}

\begin{center}
{\Large{\bf
Nonequilibrium representative ensembles for isolated quantum 
systems} \\ [5mm]

V.I. Yukalov} \\ [3mm]

{\it Bogolubov Laboratory of Theoretical Physics, \\
Joint Institute for Nuclear Research, Dubna 141980, Russia}

\end{center}

\vskip 3cm

\begin{abstract}

An isolated quantum system is considered, prepared in a nonequilibrium 
initial state. In order to uniquely define the system dynamics, one has 
to construct a representative statistical ensemble. From the principle 
of least action it follows that the role of the evolution generator is 
played by a grand Hamiltonian, but not merely by its energy part. A 
theorem is proved expressing the commutators of field operators with 
operator products through variational derivatives of these products. 
A consequence of this theorem is the equivalence of the variational 
equations for field operators with the Heisenberg equations for the 
latter. A finite quantum system cannot equilibrate in the strict sense. 
But it can tend to a quasi-stationary state characterized by ergodic 
averages and the appropriate representative ensemble depending on 
initial conditions. Microcanonical ensemble, arising in the eigenstate 
thermalization, is just a particular case of representative ensembles. 
Quasi-stationary representative ensembles are defined by the principle 
of minimal information. The latter also implies the minimization of an 
effective thermodynamic potential.  
\end{abstract}

\vskip 2cm
 

\vskip 1cm


\section{Introduction}

Physics of finite quantum systems is now attracting strong attention 
because of the widespread use of such systems in a variety of applications. 
As a few examples, it is possible to mention finite spin systems, quantum 
dots and wells, trapped atoms and ions, and numerous nano- and mesoscopic 
devices in quantum electronics. Finite systems can be well isolated from 
surrounding, which poses several principal questions on the behaviour of 
isolated quantum systems. Such isolated systems possess the properties 
that can be essentially different from those of bulk matter typical of 
condensed-matter materials. Especially difficult is the problem of 
describing nonequilibrium behaviour of finite quantum systems. Different 
sides of this problem have been surveyed in reviews [1-8]. One usually 
studies nonequilibrium effects in particular systems, such as spin 
assemblies [5,9-18], trapped atoms [4,6,8,14-17], atoms in double wells 
[18-22], and quantum dots [23].

The aim of the present paper is to analyze several principal problems 
that are common for the dynamics of many isolated quantum systems 
prepared in a strongly nonequilibrium initial state. The basic questions 
that will be studied here are as follows: (i) What is the general rule 
for constructing nonequilibrium statistical ensembles? (ii) How the 
variational equations for field operators are connected to the Heisenberg 
equations of motion? (iii) In what sense an isolated quantum system could 
equilibrate? The specific feature of the present paper is that the answers 
to these questions are given not by treating some particular systems but 
are based on the general principles, such as the principle of least action 
and the principle of minimal information.

Throughout the paper, the system of units is employed, where the Planck 
and Boltzmann constants are set to one.

\section{Grand Hamiltonian as evolution generator}

The behaviour of a nonequilibrium quantum system is characterized by 
a nonequilibrium quantum ensemble that, by definition, is the pair 
$\{\cF,\hat\rho(t)\}$ of the quantum state of microstates $\cF$ and
a statistical operator $\hat{\rho}(t)$ parametrized by time $t$. The 
temporal evolution of the statistical operator is prescribed by the 
equality
\be
\label{1}
 \hat\rho(t) = \hat U(t) \hat\rho(0) \hat U^+(t) \; ,
\ee
in which the evolution operator $\hat{U}(t)$ satisfies the Schr\"odinger 
equation
\be
\label{2}
 i\; \frac{d}{dt} \; \hat U(t) = H \hat U(t)\; ,
\ee
with a Hamiltonian $H$. This ensemble allows one to get the evolution 
equations for any observable quantity associated with an operator 
$\hat{A}$ from an algebra of local observables acting on $\cal{F}$. 
The measurable quantity is the average
\be
\label{3}
 \lgl \; \hat A(t) \; \rgl \equiv {\rm Tr} \hat\rho(t) \hat A =
{\rm Tr} \hat\rho \hat A(t) \; ,
\ee
where $\hat{A}\equiv \hat{A}(0)$, $\hat{\rho}\equiv\hat{\rho}(0)$, the 
trace is over $\cF$, and 
$$
\hat A(t) \equiv \hat U^+(t) \hat A(0) \hat U(t) \;  .
$$

Since the dynamics in Eqs. (1) and (2) is generated by the Hamiltonian 
$H$, it is termed the {\it evolution generator}. Thus one comes to the 
problem of correctly defining the evolution generator, which would make 
the quantum ensemble completely described. Constructing a statistical 
ensemble, one must always keep in mind that such an ensemble has to be 
{\it representative}, that is, uniquely defining the considered system 
[24,25]. As was stressed by Gibbs [26,27], for uniquely characterizing 
a statistical system, it may be necessary to define not merely an energy 
operator, but also all those additional constraints that allow for the 
unique system description. Let $\hat{C}_i$ be self-adjoint operators, 
enumerated by the index $i= 1,2,\ldots$, whose average values define 
such {\it additional conditions}
\be
\label{4}
 C_i(t) = \lgl \; \hat C_i(t) \; \rgl \; .
\ee

A general principle for deriving evolution equations is the principle 
of least action, or the principle of stationary action. For quantum 
systems, an action is an operator functional
\be
\label{5}
\hat S = \int \hat L(t) \; dt \; , \qquad 
\hat L(t) = \hat E(t) - \hat H(t) \;  ,
\ee
in which $\hat L(t)$ is a Lagrangian, with $\hat E(t)$ being an energy 
operator and $\hat{H}(t)$, the energy part of a Hamiltonian. The 
stationarity of action requires that action (5) be stationary, under 
the given conditions (4). This is equivalent to the stationarity of 
the {\it effective action}
\be
\label{6}
 S_{eff} = \int \left [ \hat L(t) - \sum_i \lbd_i \hat C_i(t)
\right ] \; dt \;  ,
\ee
where $\lambda_i$ are the Lagrange multipliers guaranteeing the 
validity of constraints (4). The effective action can be rewritten as
\be
\label{7}
 S_{eff} = \int \left [ \hat E(t) - H(t) \right ] \; dt \;  ,
\ee
with the {\it grand Hamiltonian}
\be
\label{8}
H(t) \equiv \hat H(t) + \sum_i \lbd_i \hat C_i(t) \;  .
\ee
In this way, to be uniquely defined, preserving all required constraints, 
the system evolution has to be generated by the grand Hamiltonian (8), 
but not merely by the energy part $\hat{H}$. 

Let the quantum system be characterized by field operators $\psi(x,t)$, 
in which $x$ is a set of the variables describing the system. Then the 
stationarity of action (7) implies the equation
$$
\dlt S_{eff} = \frac{\dlt S_{eff}}{\dlt\psi(x,t)} \; \dlt \psi(x,t)
+ \frac{\dlt S_{eff}}{\dlt\psi^\dgr(x,t)} \; \dlt\psi^\dgr(x,t) = 0\; .
$$
With the energy operator
\be
\label{9}
 \hat E(t) = \int \psi^\dgr(x,t) \; i \; \frac{\prt}{\prt t} \;
\psi(x,t) \; dx \; ,
\ee
one gets the {\it variational equation}
\be
\label{10}
 i\; \frac{\prt}{\prt t} \; \psi(x,t) = 
\frac{\dlt H(t)}{\dlt\psi^\dgr(x,t)} \;  ,
\ee
plus its Hermitian conjugate, playing the role of the Euler-Lagrange 
equations for quantum systems.

It is important to stress it again that the evolution generator is the 
grand Hamiltonian (8). Only then the system behavior can be uniquely 
defined. The internal terms of Hamiltonian (8) do not need to be mutually 
commutative. That is, the condition operators $\hat{C}_i$ do not have to 
commute with the energy part $\hat{H}$. The whole evolution generator $H$, 
of course, commutes with itself, hence $H(t)=H$ is an integral of motion.

\section{Relation between variational derivatives and commutators}

For a quantum system, characterized by field operators, the equations 
of motion are usually written in the Heisenberg form, where, instead of 
the variational derivative in Eq. (10), one has a commutator of the 
field operator with $H$. One tacitely assumes that these forms should 
be equivalent, though, to the knowledge of the author, no general proof 
of this has been given. Here, we prove a very general relation between 
variational derivatives and commutators, from which the equivalence of 
the Euler-Lagrange type variational equation (10) and the Heisenberg 
equation for the field operators follows as a particular case.

The field operators, depending on the particle statistics, satisfy 
either commutation or anticommutation relations
\be
\label{11}
 \left [ \psi(x,t) , \; \psi^\dgr(x',t) \right ]_\mp = 
\dlt(x-x') \; , \qquad 
\left [ \psi(x,t) , \; \psi(x',t) \right ]_\mp = 0 \; ,
\ee
in which the upper sign is for Bose and lower, for Fermi statistics. 
Let the operator
\be
\label{12}
\hat P_{mn}(t) \equiv P_m^+(t) P_n'(t)
\ee
be a product of two terms
\be
\label{13}
 P_m^+(t) \equiv \prod_{i=1}^m \psi^\dgr(x_i,t) \;  , \qquad
P'_n(t) \equiv \prod_{j=1}^n \psi(x_j',t) \;  ,
\ee
where $m$ and $n$ are any integers. Then the following proposition is 
valid.

\vskip 2mm

{\bf Theorem 1}. For the field operators, satisfying relations (11), 
the equality is valid:
\be
\label{14}
 \left [ \psi(x,t) , \; \hat P_{mn}(t) \right ] =
\frac{\dlt\hat P_{mn}(t)}{\dlt\psi^\dgr(x,t)} +
\left [ (\pm 1)^{m+n} - 1 \right ] \hat P_{mn}(t) \psi(x,t) \;  .
\ee

\vskip 2mm

{\it Proof}. For the left-hand side of Eq. (14), we have
\be
\label{15}
 \left [ \psi(x,t) , \; \hat P_{mn}(t) \right ] = 
 \left [ \psi(x,t) , \; P_{m}^+(t) \right ] P_n'(t) +
P^+_m(t)  \left [ \psi(x,t) , \; P'_{n}(t) \right ] \; .
\ee
Here for the first commutator in the right-hand side we find
$$
 \left [ \psi(x,t) , \; P^+_{m}(t) \right ] = \left [ 
(\pm 1)^m - 1 \right ] P_m^+(t) \psi(x,t) 
 + \sum_{i=1}^m (\pm 1)^{i+1} \dlt(x-x_i) 
\prod_{j(\neq i)}^m \psi^\dgr(x_j,t) \;  .
$$
By definition (12), it is clear that
$$
 \frac{\dlt\hat P_{mn}(t)}{\dlt\psi^\dgr(x,t)} =
\frac{\dlt P^+_m(t)}{\dlt\psi^\dgr(x,t)} \; P'_n(t) \;  .
$$
Taking the variational derivative, we get
$$
 \frac{\dlt P^+_{m}(t)}{\dlt\psi^\dgr(x,t)} = 
\sum_{i=1}^m (\pm 1)^{i+1} \dlt(x-x_i) 
\prod_{j(\neq i)}^m \psi^\dgr(x_j,t) \; .
$$
Comparing this with the above commutator gives
\be
\label{16}
 \left [ \psi(x,t) , \; P^+_{m}(t) \right ] =
\frac{\dlt P^+_{m}(t)}{\dlt\psi^\dgr(x,t)} +
\left [ (\pm 1)^m - 1 \right ] P_m^+(t) \psi(x,t) \; .
\ee
Direct calculations also result in
\be
\label{17}
  \left [ \psi(x,t) , \; P'_{n}(t) \right ] = 
 \left [ (\pm 1)^n - 1 \right ]  P'_{n}(t) \psi(x,t) \; .
\ee
Combining Eqs. (15) to (17), we come to relation (14). 
  
\vskip 2mm

An important consequence from this theorem follows for the operators 
from the algebra of local observables
\be
\label{18}
  \cA \equiv \left \{ \hat A(t) \right \} \; ,
\ee
which are self-adjoint operators possessing the general representation
$$
\hat A(t) = \sum_{mn} \; \frac{1}{\sqrt{m! n!} } \; \int
A_{mn}(x_1,\ldots,x_m;x_1',\ldots,x_n') \times
$$
\be
\label{19}
 \times \hat P_{mn}(x_1,\ldots,x_m;x_1',\ldots,x_n')\; dx_1\ldots dx_m 
dx_1' \ldots dx_n' \; ,
\ee
where for Bose statistics $m$ and $n$ in the summation are arbitrary, 
but for Fermi statistics the summation includes only such $m$ and $n$ 
for which $m+n$ is even. The latter restriction takes into account the 
conservation of half-integer spins of fermions. The quantity $A_{mn}$ 
does not depend on the field operators. The form of $\hat{P}_{mn}$ here 
is the same as in Eq. (12), but with explicitly shown variables.

\vskip 2mm

{\bf Theorem 2}. For any operator $\hat{A}(t)$ from the algebra of local 
observables (18), there exists the relation 
\be  
\label{20}
 \left [ \psi(x,t) , \; \hat A(t) \right ] =  
\frac{\dlt\hat A(t)}{\dlt\psi^\dgr(x,t)} \; .
\ee 

\vskip 2mm

{\it Proof}. Relation (20) is a straightforward consequence of equality 
(14) for the operators from the algebra of local observables (18).

\vskip 2mm

The system Hamiltonian $H$ pertains to the algebra of local observables 
(18), hence the right-hand side of the evolution equation (10) coincides 
with the commutator, according to Eq. (20). That is, the variational 
equation (10) is equivalent to the Heisenberg equation of motion.

\section{Representative ensembles for ergodic averages}

The observable quantities (3) can be conveniently rewritten invoking the
Hamiltonian basis $\{|n\rgl\}$ defined by the eigenproblem
\be
\label{21}
 H\; | \; n \; \rgl =  E_n\; | \; n \; \rgl \; .
\ee
Then the evolution of observables (3), for an isolated quantum system, 
is given by the equation
\be
\label{22}
  \lgl \; \hat A(t) \; \rgl = \sum_{mn} \rho_{mn}(t) A_{nm} \; ,
\ee
where the notation $A_{mn}\equiv\lgl m|\hat{A}|n\rgl$ for the matrix 
elements is used,
\be
\label{23}
  \rho_{mn}(t) = \rho_{mn}(0) \exp ( - i\om_{mn} t ) \; ,
\ee
and with $\omega_{mn} \equiv E_m - E_n$ being a transition frequency. 
The normalization condition for the statistical operator becomes
\be
\label{24}
 {\rm Tr} \hat\rho(t) = \sum_n \rho_{nn}(t) = 1 \;  .
\ee
As is evident, the observable quantity (22) is a quasi-periodic function 
of time. Therefore the limit of observable (22) for $t$ tending to 
infinity does not exist, and any initial state will be reproduced after 
the recurrence time that can be estimated [6] as
\be
\label{25}
  t_{rec} = \frac{2\pi}{\ep N} \; e^N \; ,
\ee
where $\varepsilon$ is a typical energy per particle. This means that 
an isolated quantum system cannot equilibrate in the strict sense.

But it may happen that, after the system has been prepared in a 
nonequilibrium initial state, the observable relaxes close to the 
ergodic average
\be
\label{26}
 \overline{ \lgl \; \hat A(t) \; \rgl } \equiv 
\lim_{\tau\ra\infty} \; \frac{1}{\tau} \; \int_0^\tau 
\lgl \; \hat A(t) \; \rgl dt
\ee
and stays close to it most of the time, except rare large deviations. 
Assuming that the spectrum $E_n$ is nondegenerate yields
\be
\label{27}
  \overline{ \lgl \; \hat A(t) \; \rgl } = 
\sum_n \rho_{nn}(0) A_{nn} \;  .
\ee
In that case, one can say that the system {\it quasi-equilibrates} 
or that it relaxes to a {\it quasi-equilibrium state} characterized 
by the ergodic average (27).

Of course, since the ergodic average does not describe a strictly 
equilibrium state, a finite quantum system may demonstrate large and 
frequent fluctuations [28]. Then it is not possible to claim that it 
relaxes to a quasi-equilibrium state. For instance, exactly solvable 
systems can demonstrate perpetual pulsation for some observables, 
though showing a kind of relaxation for others [29-31]. Generally, 
finite quantum systems relax to a quasi-equilibrium state when they 
are nonintegrable [2-8,32]. 

As is seen from Eq. (27), the ergodic average depends on the 
initially prepared state through the value $\rho_{nn}(0)$. If the 
state, which the system has relaxed to, bears information on initial 
conditions, then one says that there is no thermalization, since the 
latter requires some independence from initial conditions. 

A trivial case occurs if the initial state is a pure eigenstate  
$|j\rgl$, when $\rho_{mn} = \delta_{mj} \delta_{nj}$. Then value (22) 
is strictly stationary: $\lgl\hat{A}(t)\rgl=A_{jj}$. That is, if the 
system initial state is a pure Hamiltonian eigenstate, then there is 
no any dynamics at all. It is reasonable to assume that, if the initial 
state is characterized by a weak spread around a fixed pure state, and 
the system is not integrable, then it would quickly relax to an ergodic 
average. To concretize this, let us consider a narrow energy shell
\be
\label{28}
\mathbb{E}_j \equiv \{ E_n: \; | E_n - E_j | \leq \Dlt E_j \}
\ee
of energies deviating from a given $E_j$ not more then by $\Delta E_j$. 
And suppose that $\rho_{nn}(0)$ is nonzero only when the related energy 
$E_n$ is inside shell (28). Assume that this energy shell is so narrow 
that the matrix elements of the operators of local observables vary a 
little inside the shell, so that
\be
\label{29}
 \left | \frac{\Dlt A_j}{A_{jj} } \right | \ll 1 \; , \qquad
\Dlt A_j \equiv \max_{E_n\in\mathbb{E}_j} A_{nn} -
\min_{E_n\in\mathbb{E}_j} A_{nn} \;  .
\ee
Then, by the mean value theorem, 
$$
 \sum_{E_n\in\mathbb{E}_j} \rho_{nn}(0) A_{nn} \simeq 
A_{jj} \sum_{E_n\in\mathbb{E}_j} \rho_{nn}(0) = A_{jj} \; .
$$
The latter equality takes place for any $\rho_{nn}(0)$ normalized 
according to condition (24). If so, for simplicity, one can accept the 
uniform expression
\be
\label{30}
 \rho_{nn}(0) = \frac{1}{Z_j} \; , \qquad
Z_j \equiv \sum_{E_n\in\mathbb{E}_j} 1 \;  ,
\ee   
valid inside shell (28). Summarizing, one has
\be
\label{31}
 A_{jj} \simeq   \sum_{E_n\in\mathbb{E}_j} \rho_{nn}(0) A_{nn} 
\simeq \frac{1}{Z_j} \sum_{E_n\in\mathbb{E}_j} A_{nn} \; .
\ee

Since expression (30) corresponds to a microcanonical distribution, 
one often terms the resulting Eqs. (31), the eigenstate thermalization 
hypothesis. However, this is not a hypothesis, as far as Eqs. (31) 
immediately follow from the assumed conditions (29). What should be 
called the eigenstate thermalization hypothesis [33,34] would be the 
assumption that, under the chosen conditions (29), the observable (22) 
would tend to the ergodic average (27) with the microcanonical 
distribution (30). This, however, can happen only for nonintegrable 
systems and for sufficiently narrow energy shell (28) around the energy 
of an initial stationary state.      

More generally, the initial value $\rho_{nn}(0)$ can be characterized 
by a representative ensemble uniquely defining the considered system. 
For this purpose, one can invoke the maximization of entropy under the 
given additional constraints [35-37], which is equivalent to the 
minimization of the information functional [38,39]. Formulating the 
additional constraints as in Eq. (4), with explicitly separating the 
condition for the internal energy $E$, we have the information 
functional
\be
\label{32}
I [ \hat\rho ] = {\rm Tr} \hat\rho \ln \hat\rho + \lbd_0
\left ( {\rm Tr} \hat\rho - 1 \right ) +
\bt \left ( {\rm Tr}\hat\rho \hat H - E \right ) + 
\bt \sum_i \lbd_i \left ( {\rm Tr}\hat\rho \hat C_i - 
C_i \right ) \; ,
\ee
where $\lambda_0, \beta$, and $\lambda_i$ are the Lagrange multipliers. 
Minimizing the information functional (32) yields
\be
\label{33}
 \hat\rho = \frac{1}{Z} \; e^{-\bt H} \; , \qquad
Z \equiv {\rm Tr} e^{-\bt H} \;  ,
\ee
with the same grand Hamiltonian (8) that generates the system dynamics. 
Due to expression (33), we have 
\be
\label{34}
 \rho_{nn}(0) = \frac{1}{Z} \; e^{-\bt E_n} \;  .
\ee
The Lagrange multipliers $\beta = \beta(E, C_1, C_2, \ldots)$ and 
$\lambda_i = \lambda_i(E, C_1, C_2, \ldots)$ are defined through the given 
values of observables $E$ and $C_i$ by the equations
$$
 {\rm Tr} \hat\rho \hat H = E \; , \qquad 
{\rm Tr} \hat\rho \hat C_i = C_i \;  .
$$
The microcanonical distribution (30) is a particular case of distribution 
(34) under $\beta \rightarrow 0$.

The pair $\{\cal{F}, \hat{\rho}\}$, with the statistical operator (33), 
forms a representative ensemble describing the quasi-equilibrium state 
which a finite quantum system equilibrates to. As is evident, this ensemble 
is defined by invoking the information on the diagonal elements of 
$\rho_{nn}(0)$ at the initial moment of time. As has been stressed above, 
the necessity of taking into account all constraints uniquely defining the 
system, composing a grand Hamiltonian (8), was emphasized by Gibbs [26,27]. 
The term {\it representative ensemble} was used by Tolman [24] and ter Haar 
[25]. The derivation of the statistical operator by maximizing entropy, 
under the given constraints, was advanced by Shannon [35,36] and later 
advocated by Janes [37]. One sometimes calls the representative ensembles 
as generalized, or conditional, Gibbs ensembles. But one should not forget 
that such grand ensembles were introduced by Gibbs. Representative ensembles 
can be defined for stationary as well as nonstationary systems [38-43].
More details and an extensive list of references can be found in review [6].  

If the statistical operator (33) depends on some unspecified parameters or
functions, these can be specified by minimizing the information functional 
(32). Substituting expression (33) into functional (32) gives
\be
\label{35}
 I [ \hat\rho ] = \bt ( \Om - E ) \; , \qquad 
\Om \equiv - T \ln Z \;  ,
\ee
where $\Omega$ is the grand thermodynamic potential. Since the value 
$E$ here is fixed, the minimization of the information functional is 
equivalent to the minimization of the thermodynamic potential:
$$
 \min I [ \hat\rho ] \; \longleftrightarrow \; \min \Om \;  .
$$
   
In this way, when a finite quantum system equilibrates to a 
quasi-equilibrium state, the latter is described by a representative 
ensemble, whose definition is based on the information on the initially 
prepared state. Of special importance is the prescription of the initial 
state symmetry when the dynamics of a finite system involves the creation 
of topological defects [4,44,45].

\section{Conclusion}

The evolution of a finite quantum system is considered, starting from 
a nonequilibrium initial state. Being based on the principle of least 
action, it is shown that the system evolution is generated by a grand 
Hamiltonian. This defines a nonequilibrium representative ensemble, 
taking into account all constraints required for the unique system 
description.   

A theorem is proved expressing the commutators of field operators 
with the operator products through variational derivatives of the 
latter. Using this theorem, it is proved that the variation of the 
operators from the algebra of local observables is related to the 
commutators of these observables with field operators. The variational 
evolution equations, playing the role of the quantum Euler-Lagrange 
equations, are shown to be equivalent to the Heisenberg equations of 
motion.

A finite quantum system cannot equilibrate in the strict sense. But 
it can tend to a quasi-stationary state corresponding to the ergodic 
averages. This state exists in the time interval between the relaxation 
time and the recurrence time. The resulting quasi-stationary states are 
characterized by representative ensembles, whose definition involves 
information on the initially prepared state of the system. Microcanonical 
ensemble is a particular case of representative ensembles. Generally, 
such representative ensembles are defined by minimizing the appropriate 
information functionals. The minimum of the latter also implies the 
minimum of the related thermodynamic potential.     

Strictly speaking, equilibration requires that all system observables 
would relax to their quasi-stationary values. Relaxation times for 
different observables can be different. Integrable systems can display 
relaxation for some observables but the absence of such a relaxation 
for others. Nonintegrable systems always relax to quasi-stationary 
states described by representative quantum ensembles.

\vskip 5mm
{\bf Acknowledgement}

\vskip 3mm
Financial support from the Russian Foundation for Basic Research 
is acknowledged. Useful discussions and the help of E.P. Yukalova 
are appreciated.

\end{document}